\documentclass[article,twocolumn,superscriptaddress]{revtex4-1}
\usepackage{graphicx}
\usepackage{dcolumn}
\usepackage{bm}
\usepackage{amssymb, amsmath, amsthm}	

\begin{document}

\title{Spin-orbit dynamics of single acceptor atoms in silicon}
\author{J. van der Heijden}
\email{j.vanderheijden@unsw.edu.au; s.rogge@unsw.edu.au}
\affiliation{School of Physics and Australian Centre of Excellence for Quantum Computation and Communication Technology, UNSW, Sydney, Australia}
\author{T. Kobayashi}
\affiliation{School of Physics and Australian Centre of Excellence for Quantum Computation and Communication Technology, UNSW, Sydney, Australia}
\author{M.G. House}
\affiliation{School of Physics and Australian Centre of Excellence for Quantum Computation and Communication Technology, UNSW, Sydney, Australia}
\author{J. Salfi}
\affiliation{School of Physics and Australian Centre of Excellence for Quantum Computation and Communication Technology, UNSW, Sydney, Australia}
\author{S. Barraud}
\affiliation{University of Grenoble-Alpes and CEA, LETI, MINATEC, 38000 Grenoble, France}
\author{R. Lavieville}
\affiliation{University of Grenoble-Alpes and CEA, LETI, MINATEC, 38000 Grenoble, France}
\author{M.Y. Simmons}
\affiliation{School of Physics and Australian Centre of Excellence for Quantum Computation and Communication Technology, UNSW, Sydney, Australia}
\author{S. Rogge}
\email{j.vanderheijden@unsw.edu.au; s.rogge@unsw.edu.au}
\affiliation{School of Physics and Australian Centre of Excellence for Quantum Computation and Communication Technology, UNSW, Sydney, Australia}

\begin{abstract}
Two-level quantum systems with strong spin-orbit coupling allow for all-electrical qubit control and long-distance qubit coupling via microwave and phonon cavities, making them of particular interest for scalable quantum information technologies. In silicon, a strong spin-orbit coupling exists within the spin-3/2 system of acceptor atoms and their energy levels and properties are expected to be highly tunable. Here we show the influence of local symmetry tuning on the acceptor spin-dynamics, measured in the single-atom regime. Spin-selective tunneling between two coupled boron atoms in a commercial CMOS transistor is utilised for spin-readout, which allows for the probing of the two-hole spin relaxation mechanisms. A relaxation-hotspot is measured and explained by the mixing of acceptor heavy and light hole states. Furthermore, excited state spectroscopy indicates a magnetic field controlled rotation of the quantization axes of the atoms. These observations demonstrate the tunability of the spin-orbit states and dynamics of this spin-3/2 system.
\end{abstract}

\maketitle

Single impurities in semiconductors offer an extensive variety of spin systems~\cite{Koenraad:2011, Zwanenburg:2013}, desirable for the emerging fields of quantum technologies, such as quantum communication~\cite{Kurtsiefer:2000}, quantum sensing~\cite{Tetienne:2014, Bonato:2016}, and quantum information~\cite{Muhonen:2014, Smolenski:2016}. Impurities with spin-1/2, such as phosphorus in silicon, are promising two-level spin systems for quantum technologies~\cite{Kane:1998}. Alternatively, impurities with spin-3/2 offer the possibility to isolate a two-level system whose quantum properties strongly depend on the specific splitting and mixing of the spin-3/2 states, thereby providing a highly tunable quantum system. Several spin-3/2 impurities have been investigated in recent studies, such as single Co atoms~\cite{Otte:2008, Toskovic:2016} and Si vacancies in silicon carbide~\cite{Kraus:2014, Widmann:2015}, as well as spin-3/2 hole quantum dots~\cite{Hu:2012, Prechtel:2016, Maurand:2016}. However, the adjustability of impurity spin-3/2 systems has not yet been demonstrated. Here this tunability is shown using the until now little explored spin-3/2 system of a single acceptor in silicon, which has a level configuration and environmental couplings predicted to yield highly tunable qubits~\cite{Ruskov:2013, Salfi:2016.2, Salfi:2016.3, Abadillo-Uriel:2016}. Of particular interest is the possibility to tune the acceptor system to a regime with a large dipole coupling whilst being robust against charge noise~\cite{Salfi:2016.3}. By using a silicon CMOS transistor to access the single acceptor regime~\cite{vanderHeijden:2014}, which allows excellent control over the environment of the atom, this work demonstrates control over the level configuration of a single acceptor in silicon and probes the resulting influence on the spin-dynamics of the acceptor states.

A predicted strong spin-orbit coupling for the heavy hole ($m_j=\pm3/2$) and light hole ($m_j=\pm1/2$) states makes acceptors in silicon good candidates for spin-orbit qubits~\cite{Ruskov:2013, Salfi:2016.2, Salfi:2016.3}. This type of qubit is advantageous for the all-electrical qubit control by local gates~\cite{Nowack:2007, NadjPerge:2010, Kawakami:2014, Maurand:2016} and the long-range qubit coupling via the interaction with microwave resonators~\cite{Petersson:2012, Gustafsson:2014, Viennot:2015}. The level configuration of the acceptor atom is predicted to strongly depend on the local strain, electric field, magnetic field, and the proximity of the atom to an interface~\cite{Bir:1963.1, Bir:1963.2, Abadillo-Uriel:2016}. As a consequence the essential properties of possible two-level quantum systems created from these levels, such as the electric dipole coupling, relaxation time and coherence time, can be drastically changed~\cite{Ruskov:2013, Salfi:2016.2, Salfi:2016.3}. So far spin-dynamic experiments on acceptor ensembles in bulk silicon have been hindered by a large inhomogeneous broadening, originating from the random heavy-light hole splitting induced by each atom's local environment~\cite{Song:2011}. By addressing individual acceptor atoms, which is essential to operate them as qubits, this effect of inhomogeneous broadening is circumvented. Single acceptor measurements have demonstrated the effects of electric field~\cite{Calvet:2007.1}, local strain~\cite{Calvet:2007.2}, a nearby interface~\cite{Mol:2015}, magnetic field~\cite{vanderHeijden:2014}, and Hubbard-like interactions between two acceptors~\cite{Salfi:2016.1}. 

\begin{figure}[t]
\begin{center}
\includegraphics[width=88mm]{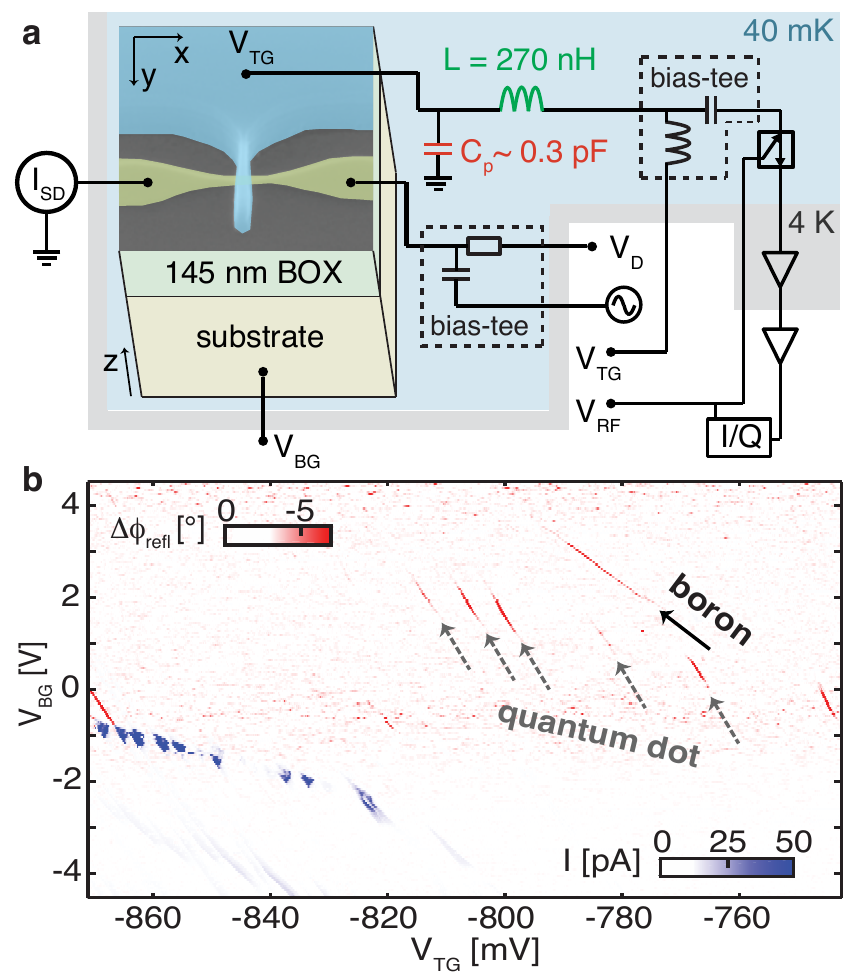}
\caption{Detection of a single acceptor atom. \textbf{a}, Scanning electron microscope image of the used tri-gate transistor in a schematic diagram of the experimental setup for transport and rf gate reflectometry measurements. \textbf{b}, Change in reflected phase from the LC-circuit (red) compared to the measured source-drain current (blue), both at 1 mV $V_\text{D}$. The signals of a boron atom (black arrow) and an unintentional quantum dot (grey arrows) are identified.}
\label{fig:Figure1}
\end{center}
\end{figure}

Here we demonstrate the ability to tune the level configuration and, as a consequence, the hole spin dynamics of an acceptor atom, thereby showing the fundamental versatility of this spin-3/2 system. For this study a two-hole system confined to two boron atoms in a state-of-the-art silicon CMOS transistor (see Supplementary information A for fabrication details) is investigated using a combination of single hole transport~\cite{vanderHeijden:2014, Li:2015, Voisin:2016} and dispersive radio-frequency (rf) gate reflectometry techniques~\cite{Petersson:2010, Colless:2013, Verduijn:2014, Gonzalez-Zalba:2015, House:2015} (see Fig.~\ref{fig:Figure1}a). The crossover point between one hole on each acceptor, the (1,1) state, and both holes on one acceptor, the (2,0) state, is used to probe the spin state of the double hole system, making use of spin selective tunneling~\cite{Johnson:2005}. Magnetic field dependent measurements of the relaxation rate and the excited state spectrum of the two-hole system show that the quantization axis of the angular momentum can be rotated with the applied field, which strongly tunes the mixing and splitting between the heavy and light hole states. The amount of control over the level configuration of a single acceptor atom demonstrated here is an essential step towards an acceptor qubit. 

\subsection*{Identification of individual acceptors}

RF gate reflectometry ~\cite{Petersson:2010, Colless:2013, Verduijn:2014, Gonzalez-Zalba:2015, House:2015} is used to detect individual acceptors. The charging of a boron atom in the nano-transistor is observed as a shift of the phase, $\Delta\phi_\text{refl}$, or the amplitude, $\Delta A_\text{refl}$, of the reflection from a LC-circuit connected to the top-gate, when driven near the resonance frequency with $V_\text{RF}$. This shift is caused by a change in admittance of the LC-circuit induced by tunneling of holes when a localized site in the transistor changes its charge occupation~\cite{Verduijn:2014}. In a map of topgate voltage ($V_\text{TG}$) and backgate voltage ($V_\text{BG}$) in the presence of a small drain voltage ($V_\text{D}$ = 1 mV), shown in Fig.~\ref{fig:Figure1}b, some charging lines appear far below the onset of hole transport current $I_\text{SD}$. These lines are identified as the charging of localized sites which are only strongly tunnel coupled to one of the electrodes. The slopes of these lines correspond to the capacitive coupling ratio to the top and back gate. Multiple lines with equal slopes (grey arrows in Fig.~\ref{fig:Figure1}b) imply the charging of several charge states of the same localized state, here likely to be an unintentional quantum dot at the Si/SiO$_2$ interface~\cite{Voisin:2014}. In contrast, a boron atom can bind no more than two holes~\cite{Aleksandrov:1975}. A single charging line, unaccompanied with any counterparts of equal slope (black arrow in Fig.~\ref{fig:Figure1}b), indicates the charging of the first hole onto a single boron atom. The location of this atom is estimated by comparing the capacitive couplings to the four transistor electrodes (see Supplementary information B), which reveal a position not directly beneath the top gate, but instead close to the drain lead. This verifies that this charging line does not come from the charging of a gate-defined quantum dot.

\begin{figure*}
\begin{center}
\includegraphics[width=183mm]{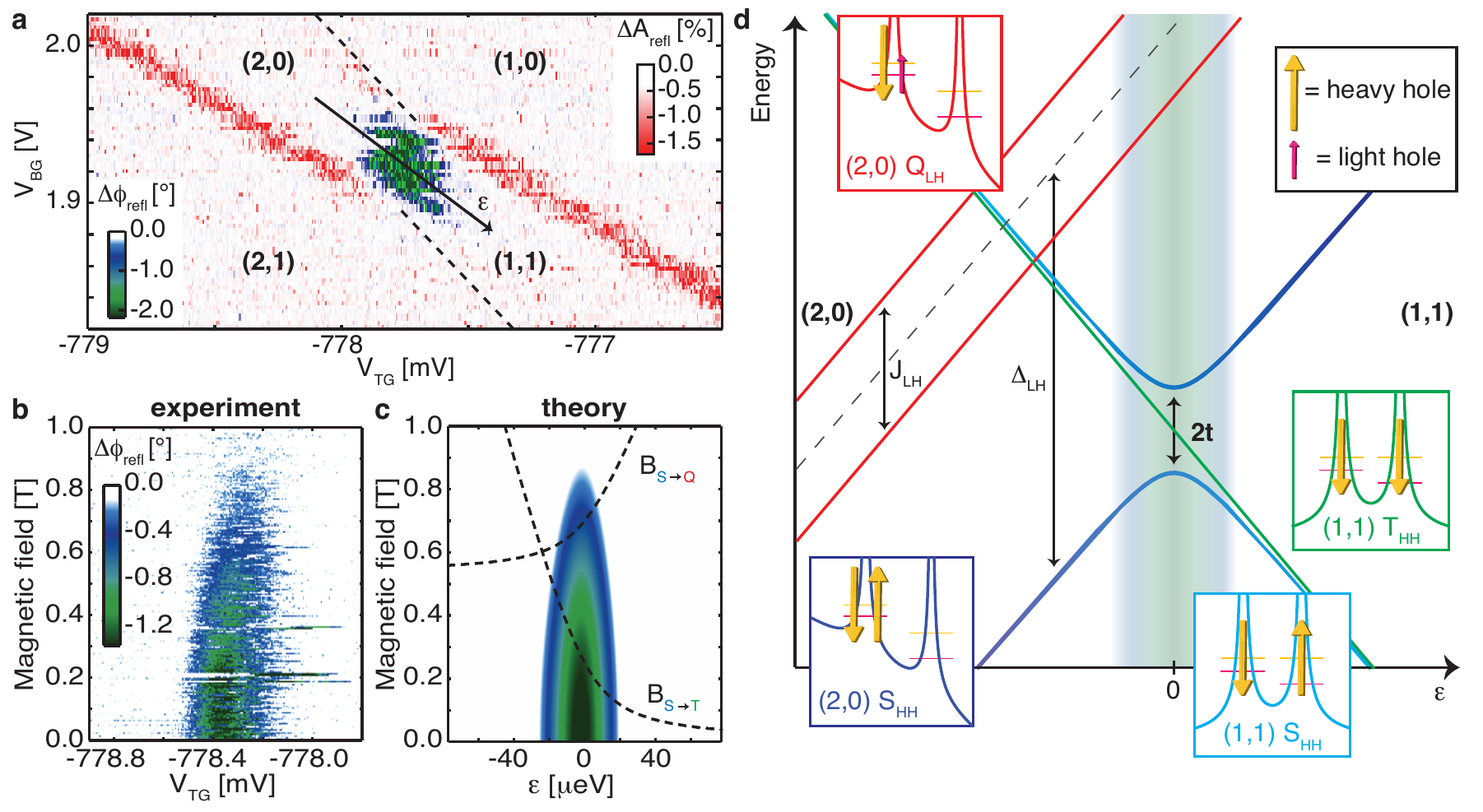}
\caption{Two-hole system. \textbf{a}, Reflected amplitude and phase response showing a break in the acceptor-lead charging line (red), caused by the charging of a nearby acceptor atom (dashed lines). \textbf{b}, \textbf{c}, Measured and calculated magnetic field dependence of the inter-acceptor tunnelling phase response. The magnetic field yielding the singlet-triplet crossing $B_{\text{S}\rightarrow\text{T}}$ and singlet-quadruplet crossing $B_{\text{S}\rightarrow\text{Q}}$ are plotted in \textbf{c} as a function of detuning by dashed curves. \textbf{d}, Schematic overview of the relevant two-hole states. The (1,1) and (2,0) $S_\text{HH}$ states [blue], (1,1) T$_\text{HH}$ states [green], and (2,0) Q$_\text{LH}$ states [red] are shown. While this subset of two-hole states is sufficient to describe our experiments, a full description of the spin-orbit states can be found in Supplemetary information D. Insets show a depiction of the potential landscape and possible spins involved in these states.}
\label{fig:Figure2}
\end{center}
\end{figure*}

A sudden electrostatic shift in the charging signal of this acceptor, shown in Fig.~\ref{fig:Figure2}a, is the signature of the charging of another localized site nearby. Further measurements of this state, presented later in this paper, show that this site is a second boron atom. The negative $\Delta\phi_\text{refl}$ found within the break of the line indicates the tunneling of holes between the two atoms. A detailed study of the reflected signal in the frequency domain shows that the acceptor-lead tunnel rate is comparable to the resonant frequency (583 MHz) of the LC-circuit (Supplementary information C). In contrast, the inter-acceptor tunnel rate is found to be much faster than this resonant frequency. The tunneling of holes from either lead to the second acceptor atom is not observed, likely due to tunnel rates much slower than the resonant frequency. 

\subsection*{Two-hole system}

The magnetic field dependence of the inter-acceptor tunneling, shown in Fig.~\ref{fig:Figure2}b, reveals the number of holes bound to the acceptor atoms. We observe the signature of Pauli spin blockade in the reflected signal~\cite{House:2015, Betz:2015} when a magnetic field is applied along the axis of the nanowire that forms the top gate of the transistor (y-direction in Fig.~\ref{fig:Figure1}a). Approximately at 0.25 T the reflectometry signal starts to weaken due to spin-selective tunneling associated with the increasing thermal population of a triplet state. This observation of spin-selective tunneling indicates the presence of an even number of holes bound to the two acceptor atoms, from which it is concluded that the inter-acceptor tunneling corresponds to the (1,1)$\leftrightarrow$(2,0) transition.

To describe the two-hole system we use a theoretical framework to construct the (1,1) and (2,0) states from single hole states (Supplementary information D). The four lowest energy states of a single hole bound to an acceptor in silicon can be described by angular momenta of $\pm3/2$ (heavy holes) and $\pm1/2$ (light holes). A distortion of the symmetry of the wave function can lift the degeneracy of heavy and light holes~\cite{Calvet:2007.1, Calvet:2007.2, Mol:2015}. Our experiments are consistent with heavy hole ground states for both acceptors, with energy splitting $\Delta^i_\text{LH}$ between the heavy and light hole states of acceptor $i$. The total energy of the two-hole system at zero magnetic field is shown as a function of the energy level detuning $\epsilon$ between the atoms in Fig.~\ref{fig:Figure2}d. To interpret our measurements it is sufficient to only consider the lowest spin manifold for the (1,1)-configuration (right side of Fig.~\ref{fig:Figure2}d), where one heavy hole resides on each acceptor, forming a singlet state, S$_\text{HH}$, and three triplet states, T$_\text{HH}^-$, T$_\text{HH}^0$, and T$_\text{HH}^+$. For the (2,0)-configuration (left side of Fig.~\ref{fig:Figure2}d), which is the doubly occupied acceptor A$^+$ state, the exchange energy between holes with the same angular momentum ($J_\text{HH}$, $J_\text{LL}$) or different angular momenta ($J_\text{LH}$=$J_\text{HL}$) are treated in the limit where $J_\text{HH}$, $J_\text{LL} > \Delta_\text{LH} >$ $J_\text{LH}$, which is the only regime consistent with our data. In analogy to the two-hole states of neutral group II acceptors~\cite{Kartheuser:1973} and of two closely spaced boron atoms~\cite{Salfi:2016.1}, the lowest energy manifold in the (2,0)-configuration consists of six states, split by $\Delta_\text{LH}$ and $J_\text{LH}$. The ground state is the heavy hole singlet, S$_\text{HH}$. Four states consisting of one heavy and one light hole, Q$_\text{LH}^{i}$, with $i$ denoting the angular momentum, are $\Delta_\text{LH}$ higher in energy and split by $J_\text{LH}$ into singlet and triplet states, as shown in Fig.~\ref{fig:Figure2}d. The sixth state is the light hole singlet state, S$_\text{LL}$, another $\Delta_\text{LH}$ higher in energy. 

\begin{figure*}
\begin{center}
\includegraphics[width=183mm]{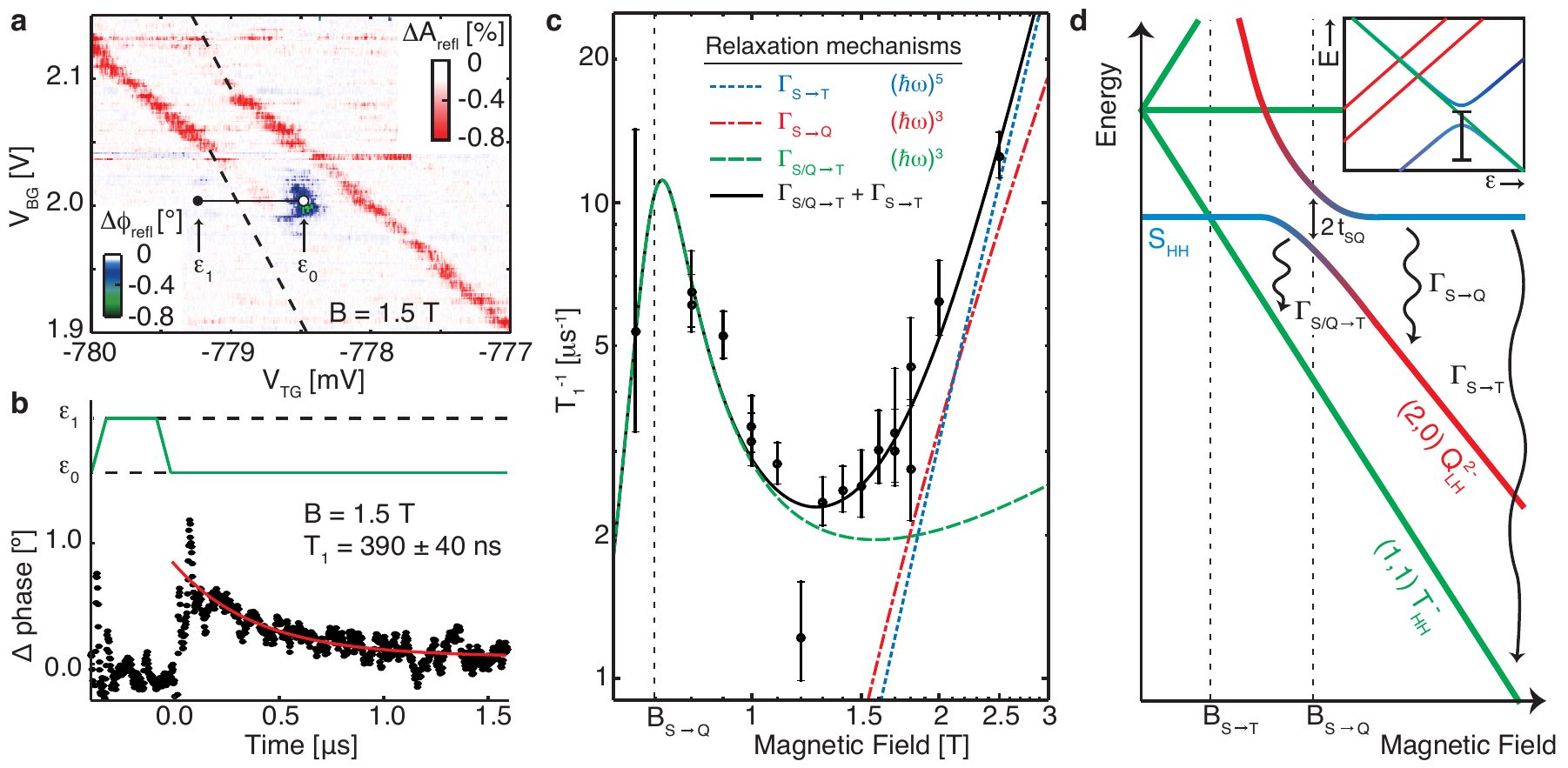}
\caption{Spin relaxation mechanisms of the two-hole system. \textbf{a}, Time averaged rf reflectometry amplitude (red) and phase (blue) measurement with continuous pulses applied to the drain, at a magnetic field of 1.5 T. The inter-acceptor tunneling is displaced from the break of the acceptor-lead tunneling signal by the Zeeman-energy of the T$_\text{HH}^-$ ground state. \textbf{b}, Relaxation time extracted from the time dependent phase signal for a pulse between the points $\epsilon_0$ and $\epsilon_1$. \textbf{c}, Magnetic field dependence of the relaxation rate T$_1^{-1}$. \textbf{d}, Schematic view of the relevant relaxation rates in this system at the zero detuning point (see inset). The data in \textbf{c} is fit to a combination of the $\Gamma_{\text{S/Q}\rightarrow \text{T}}$ rate [green line] and $\Gamma_{\text{S}\rightarrow \text{T}}$ rate [blue line] resulting in the black line. This is compared to a fit using the $\Gamma_{\text{S}\rightarrow \text{Q}}$ rate [red line] instead of the $\Gamma_{\text{S}\rightarrow \text{T}}$ rate.}
\label{fig:Figure3}
\end{center}
\end{figure*}

At zero magnetic field the tunnel coupling $t$ between the (1,1) and (2,0) S$_\text{HH}$ states gives rise to the inter-acceptor reflectometry signal, where $t$ is measured by a microwave excitation experiment to be 4.3$\pm$0.3 GHz (Supplementary information C)~\cite{Urdampilleta:2015}. At a finite magnetic field, which we label $B_{\text{S}\rightarrow\text{T}}$, the (1,1) T$_\text{HH}^-$ state becomes the ground state at the zero detuning point. Since the Pauli principle forbids tunneling between the atoms in the T$_\text{HH}^-$ state, the reflectometry becomes suppressed in this regime. Using the tunnel coupling $t$ and the thermal occupation of the two-hole states (see Supplementary information E), this magnetic field dependence of the reflectometry signal is simulated~\cite{House:2015}, as shown in Fig.~\ref{fig:Figure2}c. $B_{\text{S}\rightarrow\text{T}}$ is found to be 0.25$\pm$0.05 T, which provides an effective Land\'e g-factor ($g^*$=$gm_j$) of the T$_\text{HH}^-$ state of 1.2$\pm$0.3, using the Zeeman energy of $g^*\mu_BB$, where $\mu_B$ is the Bohr magneton. Noteworthy is that in bulk silicon the T$_\text{HH}^-$ state is expected to have $m_j$=3 and $g\sim1$, which indicates that in our experiment the $g^*$ is strongly suppressed in this low magnetic field regime. Such a situation occurs when the quantization axis of the angular momentum is not aligned with the applied magnetic field, indicating that this quantization does not come from the magnetic field but instead from either local strain, electric field, or a nearby interface.

\subsection*{Relaxation mechanisms}

A pulsed measurement technique was used to measure the spin dynamics of the two-hole system. The reflectometry signal originating from the inter-acceptor tunneling in the S$_\text{HH}$ state can be recovered at magnetic fields above $B_{\text{S}\rightarrow\text{T}}$ by pulsing the system into the (2,0) region. Such a recovered signal is indicated by a white circle in Fig.~\ref{fig:Figure3}a, where the time-averaged amplitude and phase response of the reflected signal are displayed for an applied voltage pulse of 0.8 mV to the drain for 40 out of every 200 ns and at a magnetic field of 1.5 T. This indicates a finite population of the S$_\text{HH}$ excited state at the zero detuning point, which is utilised to explore the spin-dynamics of the two-hole system using time resolved spectroscopy. Fig.~\ref{fig:Figure3}b shows the time dependence of the phase response with a voltage pulse applied between the $\epsilon_0$ and $\epsilon_1$ points, as marked in Fig.~\ref{fig:Figure3}a, and averaged over $\sim10^6$ pulse sequences. The observed exponentially decaying signal corresponds to the relaxation of the S$_\text{HH}$ state. This ability to sense the two-hole spin state could be utilized in a readout scheme for a single hole spin state, in analogy to the scheme first demonstrated by Koppens et al.~\cite{Koppens:2006}.

Fig.~\ref{fig:Figure3}c shows the relaxation rate T$_1^{-1}$ of the S$_\text{HH}$ state as a function of magnetic field. In the high magnetic field regime ($>$1.5 T), T$_1^{-1}$ increases monotonically with magnetic field. This T$_1^{-1}$ behavior is well approximated by a power-law dependence on magnetic field, which is usually expected for spin-lattice relaxation processes~\cite{Bir:1963.2, Ruskov:2013}. On the other hand, in the low magnetic field regime ($<$1.5 T), T$_1^{-1}$ reaches a maximum at $\sim$0.8 T. This implies a different spin-dynamic mechanism, likely arising from the specific interactions within this spin-3/2 acceptor system.

From the magnetic field dependence of T$_1^{-1}$ we determine what relaxation mechanisms are relevant in the probed double-acceptor system. Fig.~\ref{fig:Figure3}d illustrates the magnetic field dependence of the two-hole spin eigenenergies at zero detuning. While the (1,1) T$_\text{HH}^-$ level is the ground state at magnetic fields above $B_{\text{S}\rightarrow\text{T}}$, the (2,0) Q$_\text{LH}^{2-}$ level also falls below the S$_\text{HH}$ state above another critical magnetic field $B_{\text{S}\rightarrow\text{Q}}$. The relaxation rates from the S$_\text{HH}$ state to the T$_\text{HH}^-$ and Q$_\text{LH}^{2-}$ states, $\Gamma_{\text{S}\rightarrow\text{T}}$ and $\Gamma_{\text{S}\rightarrow\text{Q}}$, are expressed by a power law dependence on magnetic field:
\begin{equation}		
	\Gamma_{\text{S}\rightarrow \text{T[Q]}} =  A_{\text{ST[Q]}} \left(\left(B-B_{\text{S}\rightarrow\text{T[Q]}} \right)^2 + (2t_\text{ST[Q]})^2 \right)^{\gamma_\text{ST[Q]}/2}
\label{eq:EQ1}
\end{equation}
where $A_\text{ST[Q]}$ is the relaxation amplitude for the $\text{S}\rightarrow\text{T[Q]}$ process and $t_\text{ST[Q]}$ is the tunnel coupling between the S$_\text{HH}$ and the (1,1) T$_\text{HH}^-$ [(2,0) Q$_\text{LH}^{2-}$] state. While $t_\text{ST}$ is expected to be negligible, the spin-orbit interaction for light and heavy holes is expected to give a sizable $t_\text{SQ}$. Theoretical reports predict different magnetic field power laws for these relaxation process with $\gamma_\text{ST}$=5 (Kramers degenerate) and $\gamma_\text{SQ}$=3 (non-Kramers degenerate)~\cite{Ruskov:2013,Salfi:2016.3}. Both these relaxation rates monotonically increase with magnetic field like the observed T$_1^{-1}$ above $\sim$1.5 T. However, this dataset is insufficient to conclude which relaxation process is dominant in this high magnetic field range and the best fit with both the $\Gamma_{\text{S}\rightarrow\text{T}}$ and $\Gamma_{\text{S}\rightarrow\text{Q}}$ are shown in Fig.~\ref{fig:Figure3}c (Supplementary information F). 

The non-monotonic magnetic field dependence of $T_1^{-1}$ in the low magnetic field regime indicates the existence of another relaxation mechanism. The heavy-light hole mixing plays an important role in the spin relaxation around the resonance between the S$_\text{HH}$ and (2,0) Q$_\text{LH}^{2-}$ state. In the vicinity of $B_{\text{S}\rightarrow\text{Q}}$, the finite $t_\text{SQ}$ hybridises the S$_\text{HH}$ and (2,0) Q$_\text{LH}^{2-}$ states, resulting in new spin eigenstates, both containing a finite light hole component. This light hole character opens a light-to-heavy hole relaxation path to the (1,1) T$_\text{HH}^-$ state, which is expected to be much faster than the heavy-to-heavy hole relaxation process between S$_\text{HH}$ and T$_\text{HH}^-$~\cite{Ruskov:2013,Salfi:2016.3}. Since this relaxation mechanism strongly depends on the light-hole component in the hybridised states (see Supplementary information G), the relevant transition rate $\Gamma_{\text{S/Q}\rightarrow\text{T}}$ can be written as:
\begin{equation}
		\Gamma_{\text{S/Q}\rightarrow\text{T}} =A_\text{QT}\frac{2 t_\text{SQ}^2 \left(B-B_{\text{S}\rightarrow\text{T}}\right)^3}{\left(\left(B-B_{\text{S}\rightarrow\text{Q}}\right)^2+ \left(2t_\text{SQ}\right)^2\right)},
\label{eq:EQ2}
\end{equation}
where $A_\text{QT}$ is the relaxation amplitude between the Q$_\text{LH}^{2-}$ and T$_\text{HH}^−$ states, which, as mentioned, is expected to be much larger than $A_\text{ST}$. The non-monotonic magnetic field dependence of the Lorentzian peak and the large relaxation amplitude $A_\text{QT}$ give a satisfactory explanation for the observed T$_1^{-1}$ hotspot around 0.8 T.

A combination of the monotonic rate $\Gamma_{\text{S}\rightarrow\text{T}}$ and non-monotonic rate $\Gamma_{\text{S/Q}\rightarrow\text{T}}$ is used to fit the full magnetic field dependence of T$_1^{-1}$, as shown by a solid black line in Fig.~\ref{fig:Figure3}c. In this fit $B_{\text{S}\rightarrow\text{Q}}$ is found at 0.70$\pm$0.02 T, $t_\text{SQ}$ is found in magnetic field units to be 0.04$\pm$0.015 T and we extracted a ratio between the relaxation amplitudes $A_\text{QT}/A_\text{ST} \sim1.2\times10^3$.

The here observed spin-relaxation hotspot can be used as a rapid initialization of the spin-state of the acceptor atom, where the measured ratio $A_\text{QT}/A_\text{ST}$ of more than three orders of magnitude demonstrates the excellent tunability of the acceptor system. Similar hotspots, caused by the mixing of spin and orbital degrees of freedom, have been predicted and measured for systems with a strong spin-orbit coupling in III-V semiconductors~\cite{Bulaev:2005, Stano:2006, Srinivasa:2013}. Furthermore, $t_\text{SQ}$ is expected to mainly originate from the spin-orbit coupling of the heavy and light hole states. Therefore, the strong mixing between the S$_\text{HH}$ and Q$_\text{LH}^{2-}$ states, as observed in our experiment, points to the presence of a strong spin-orbit coupling for the acceptor states and thereby shows the potential to use them as electrically drivable spin-orbit qubits. 

\begin{figure*}
\begin{center}
\includegraphics[width=183mm]{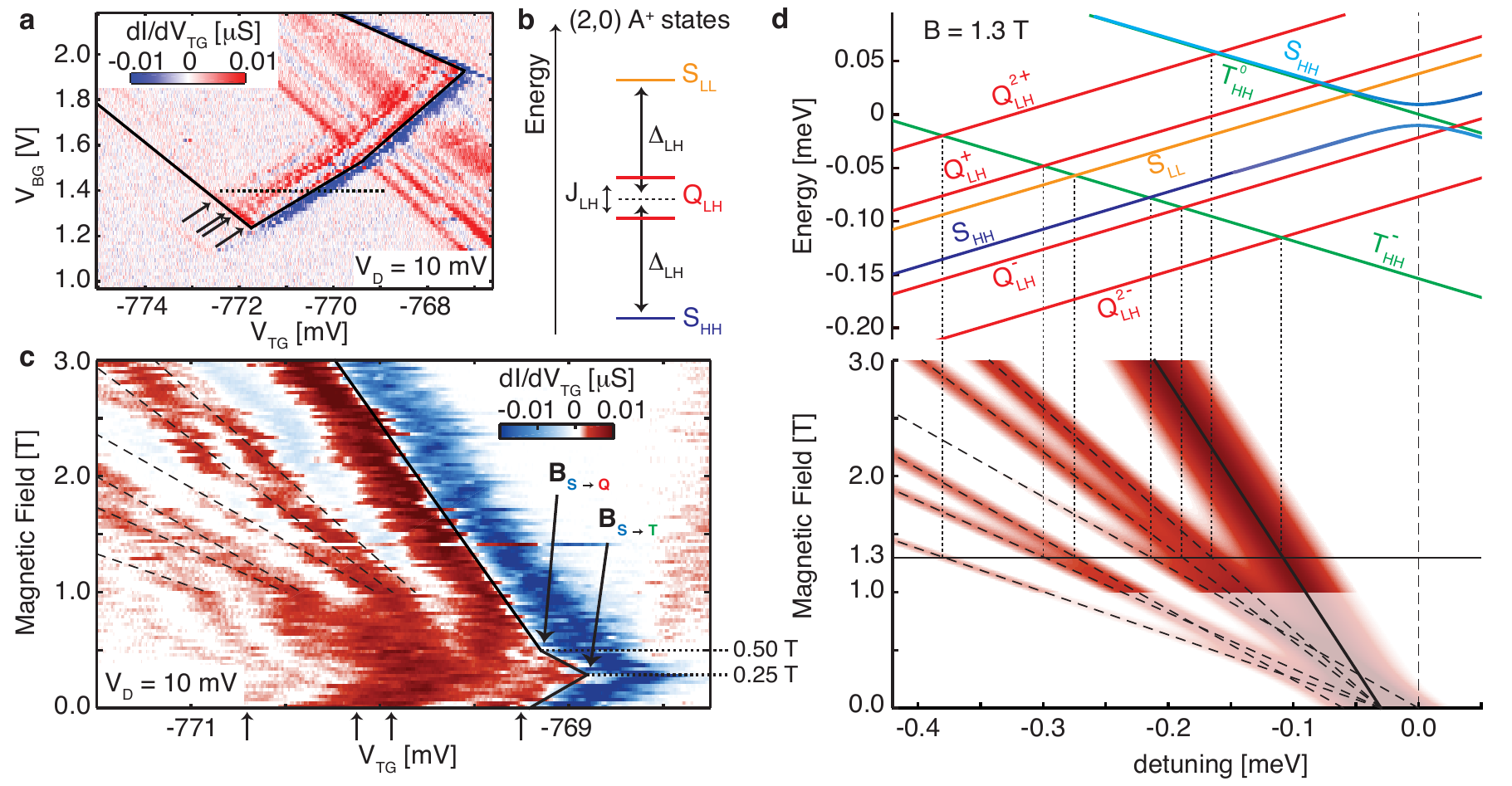}
\caption{Excited states of the two-hole system measured by transport spectroscopy. \textbf{a}, Transconductance $dI/dV_{TG}$ as a function of the top and back gate voltages at a 10 mV $V_\text{D}$. The bias triangle, marked by solid black lines, shows three excited states in addition to the ground state as indicated by arrows. \textbf{b}, The level configuration of the A$^+$ state, corresponding to the found excited states. \textbf{c}, Magnetic field dependence of the transconductance at a cut through the bias traingle at 1.4 V $V_\text{BG}$, as indicated by the dotted line in \textbf{a}. At the bottom arrows indicate the excited state spectrum as found in \textbf{a}. Two changes in the slope of the ground state at 0.25 and 0.5 T are the result of the $B_{\text{S}\rightarrow \text{T}}$ and $B_{\text{S}\rightarrow \text{Q}}$ crossings. \textbf{d}, Model of the transition energies in the high magnetic field limit. Top panel: Overview of the the (1,1)$\rightarrow$(2,0) transitions at a magnetic field of 1.3 T. Bottom panel: Simulated magnetic field behaviour of the transition energy spectrum, using the g-factors and $\Delta_\text{LH}$ extracted from the transition energies in \textbf{c} and using the same line broadening as found in this experiment.}
\label{fig:Figure4}
\end{center}
\end{figure*}

\subsection*{Two-hole excited state spectrum}

To understand the environment of the acceptor atoms we investigate the two-hole excited state spectrum. Due to their fast relaxation rate, these excited states are not accessible by the reflectometry technique and instead we use single hole transport spectroscopy~\cite{vanderHeijden:2014, Li:2015, Voisin:2016} to probe the two-hole energy spectrum. In Fig.~\ref{fig:Figure4}a the transconductance signal as a function of $V_\text{TG}$ and $V_\text{BG}$ at zero magnetic field shows a bias triangle (contour marked by a solid line), demonstrating sequential hole tunneling via both acceptors in series. Lines parallel to the triangle baseline (arrows in Fig.~\ref{fig:Figure4}a) are attributed to transitions from the (1,1) ground state to (2,0) excited states (Supplementary information D). The observed (2,0) spectrum matches the expected energy spectrum of a doubly charged acceptor state, shown in Fig.~\ref{fig:Figure4}b. This supports our identification of the second confinement site as an acceptor atom, which is further validated since our measurement is not compatible with the regular spaced excited state spectrum expected for quantum dots. The (2,0) $\Delta_\text{LH}$ and $J_\text{LH}$ are extracted as 110$\pm$10 $\mu$eV and 36$\pm$5 $\mu$eV respectively. 

The measured $\Delta_\text{LH}$ reflects the effect of local fields on the symmetry of the acceptor states. However, the direction of the resulting quantization axis for the angular momenta is unknown, which makes it impossible to predict the exact Zeeman interaction for low magnetic fields~\cite{Bir:1963.1, Bir:1963.2}. The extracted $J_\text{LH}$ reflects the mixing between light and heavy holes, associated with the spin-orbit coupling. We can compare this to $t_\text{SQ}$, taking into account that this tunnel coupling carries an additional component, as the S$_\text{HH}$ state at zero detuning is in a superposition of (1,1) and (2,0) states. Converting $t_\text{SQ}$ to an energy scale using 1.5 $\lesssim g*\lesssim$ 3, gives $t_\text{SQ}$ in the range of 3 to 7 $\mu$eV. Although somewhat lower than $J_\text{LH}$, likely caused by the rotation of the quantization axis at this magnetic field as discussed later, both these parameters show a strong spin-orbit coupling, which is crucial to envision electrical manipulations of the acceptor spin states.

Next we discuss the magnetic field dependence of the transition energy spectrum, as shown in Fig.~\ref{fig:Figure4}c, which is measured on a cut through the bias triangle at $V_\text{BG}$=1.4 V (dotted line in Fig.~\ref{fig:Figure4}a). The baseline of the triangle, indicated with a solid line in Fig.~\ref{fig:Figure4}c, changes its slope twice. The first change around 0.25 T indicates the singlet-triplet crossing at $B_{\text{S}\rightarrow\text{T}}$, in agreement with the observation of spin blockade in Fig.~\ref{fig:Figure2}b. The second change around 0.50 T indicates the change of the (2,0) ground state from S$_\text{HH}$ to Q$_\text{LH}^{2-}$, corresponding to the same anti-crossing that generates the relaxation hotspot in Fig.~\ref{fig:Figure3}b, but here occurring in the negative detuning region and therefore at a lower magnetic field. 

In the high magnetic field regime ($>$1 T) we observe well established resonance lines. The number of resolvable features is larger than at zero magnetic field, indicating spin split excited states. In this regime the Zeeman energy dominates over $\Delta_\text{LH}$ and therefore the quantization axis of the angular momenta of the two-hole states align with the external magnetic field. Ignoring any mixing terms other than the S$_\text{HH}$ mixing, the magnetic field dependence of the transition energies can be easily described by using $m_j=\pm3/2$ for heavy holes and $m_j=\pm1/2$ for light holes to calculate the Zeeman shifts, as shown in Fig.~\ref{fig:Figure4}d. Using this model, the g-factors and $\Delta_\text{LH}$ can be extracted by fitting the slopes and position of the transition lines in Fig.~\ref{fig:Figure4}c respectively. For the doubly occupied acceptor we find $g_{3/2}$ of 0.85 and $g_{1/2}$ of 1.07 and $\Delta_\text{LH}$ of 30 $\mu$eV. For the single occupied acceptor only the heavy hole states play a role in the transitions for which we extract $g_{3/2}$ of 1.05. Our observations allow for the characterization of the acceptor system in both the low and high magnetic field regimes, which shows the extent of tuning of the acceptor states.

\subsection*{Discussion}

The detected level configurations indicate a symmetry transition of the acceptor spin-3/2 system as a function of magnetic field. The g-factors found in the high magnetic field regime are close to those found for bulk boron acceptors~\cite{Kopf:1992, Stegner:2010}, which is in stark contrast to the effective g-factor found at low magnetic fields in the spin blockade experiment of Fig.~\ref{fig:Figure2}b. Furthermore the $\Delta_\text{LH}$ of the doubly occupied acceptor is about four times smaller in the high magnetic field regime than at zero magnetic field. These changes demonstrate that the quantization-axis in the high magnetic field regime differs from the quantization axis at zero magnetic field, thereby strongly influencing the g-factor and heavy-light hole splitting. A further indication of the rotating quantization axis is the decaying amplitude of the (1,1) T$_\text{HH}^-$ to (2,0) S$_\text{HH}$ transition, which completely disappears at a magnetic field around 2 T. The observation of the reduced mixing between these states is well explained by the enhancement of the spin-blockade effect when the angular momentum becomes quantized along the magnetic field. The here demonstrated possibility to change the quantization axis and strongly influence the g-factor, heavy-light hole splitting and relaxation rate of the acceptor states by controlling the atoms environment with an external field, is a fundamental requirement for reaching the optimal conditions for acceptor-based qubits~\cite{ Salfi:2016.2, Salfi:2016.3}.

In conclusion, the interactions between heavy and light hole states of individual acceptor atoms in silicon have been analysed, establishing the evaluation of the effect of local fields on the wave function symmetry and thereby on the level configurations and spin-dynamics of acceptor atoms. Mixing between light and heavy holes is found to strongly influence the spin-dynamics of the two-hole system, leading to a hotspot in the spin-relaxation rate. Furthermore it is shown that the quantization axis of the holes can be rotated by applying a magnetic field, thereby significantly tuning the g-factor and heavy-light hole splitting. We have shown access to and control over the heavy and light hole states of individual acceptors and found evidence of a strong spin-orbit coupling in this system, thereby demonstrating the fundamental principles to implement acceptor atom spin-orbit qubits.

\section*{Supplementary information}

\subsection{Experimental setup}

The silicon tri-gate transistor used in this work is fabricated using standard CMOS technology, similar to previously reported devices~\cite{Gonzalez-Zalba:2015,Voisin:2016}. The channel, defined by the top gate, is 11 nm high, 42 nm wide and 54 nm long. On both sides of this gate, 25 nm long Si$_3$N$_4$ spacers form barriers to the highly doped source and drain regions. With a background boron doping of $5\times10^{17}$/cm$^3$ in the silicon channel, about 10 boron atoms are expected under the gate and about 10 under the spacers. 

In a dilution refrigerator at a base temperature of 40 mK, hole transport and rf gate reflectometry measurements give complementary insights in the static and dynamic behaviour of the two-hole system of interest. The hole temperature was measured from quantum-dot transport at $\sim$ 300 mK. The noise floor of the transport measurement ($\sim$100 fA) allows us to detect currents with a total tunnel rate down to 500 kHz. A surface mount 270 nH inductor, attached to the top gate, forms a tank circuit together with a parasitic capacitance of $\sim$0.3 pF. Tunnel events in the transistor with a tunnel rate higher than or equal to the resonant frequency of 583 MHz result in a change of the reflected amplitude and/or phase of this resonator. Furthermore, low intensity light was directed to the silicon chip through an optical fiber to enable the use of the lower doped substrate as a back gate.

\subsection{Location of the acceptors}

In order to estimate the location of the two atoms, we have extracted the relative capacitive coupling to the different electrodes. The hole temperature of $\sim$300 mK gives $k_{b}T \gg \hbar\Gamma$ for the acceptor-lead tunnel rate, thereby justifying a thermally broadened fit~\cite{House:2015}. The reflectometry line width shows a relative coupling to the top gate ($\alpha_\text{TG}$) and drain ($\alpha_\text{D}$) of 0.50$\pm$0.01 eV/V and 0.41$\pm$0.02 eV/V respectively, thereby placing this atom roughly halfway between these two electrodes, under the $Si_3N_4$ spacer. This relatively weak coupling to the top gate supports our identification of the acceptor atom, as gate-induced quantum dots are known to have gate couplings above $\sim$0.7 eV/V~\cite{Gonzalez-Zalba:2015, Voisin:2014}. The second boron atom has to be located closer to the centre of the channel, as the tunnelling to both electrodes is too slow to be detected.

\subsection{Estimation of tunnel rates}

\begin{figure}
\begin{center}
\includegraphics[width=88mm]{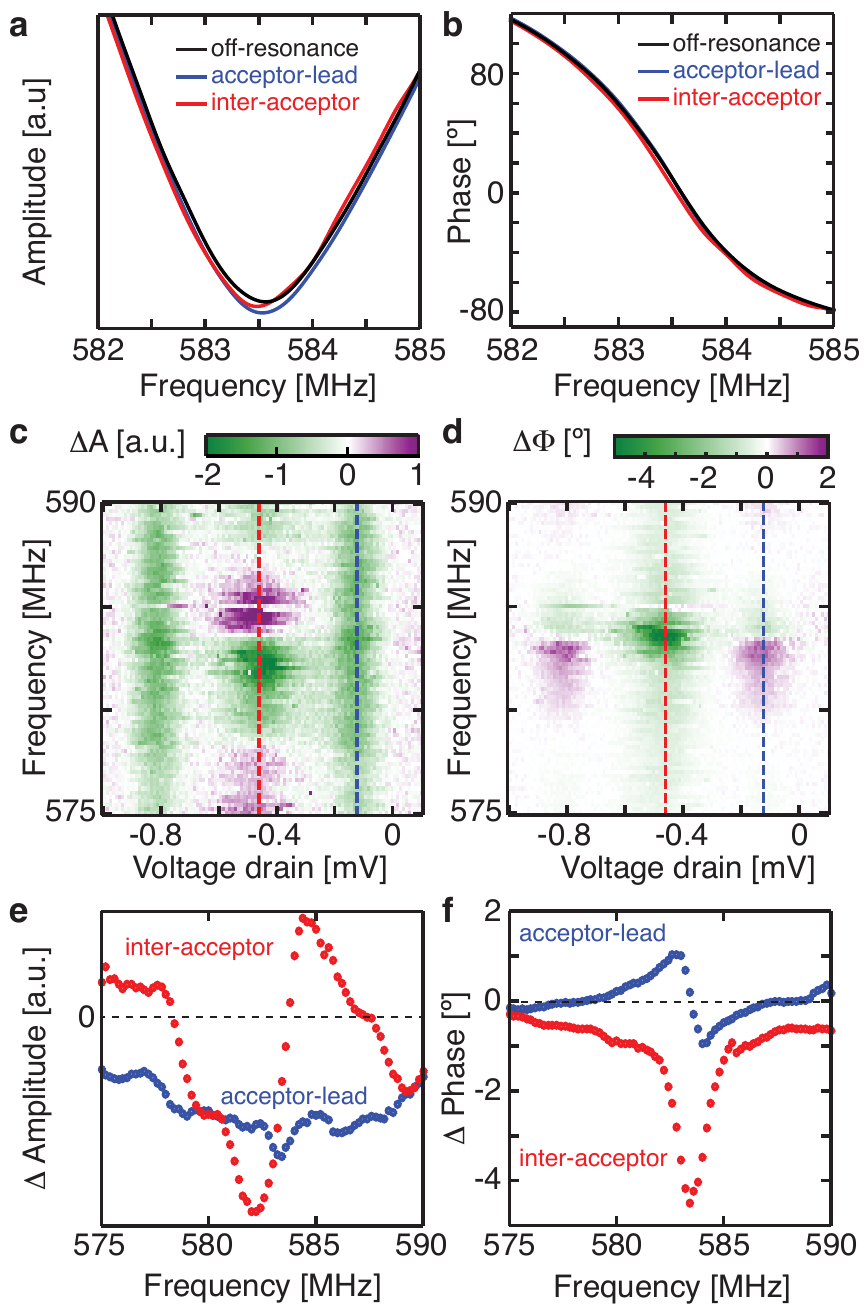}
\caption{Frequency domain analysis of the reflectometry signal. \textbf{a}, (\textbf{b},) Amplitude (phase) response of the tank circuit as a function of frequency. \textbf{c}, (\textbf{d},) Amplitude (phase) shift relative to the off-resonant signal as a function of frequency on a line cut through the (1,1)-(2,0) transition, with $V_{BG}$ = 1.93 V and $V_{TG}$ = -777 mV. \textbf{e}, (\textbf{f}) Amplitude (phase) shift for the linecuts on the red (inter-acceptor singal) and blue (acceptor-lead singal) dotted lines in \textbf{c} (\textbf{d}).}
\label{fig:FigureS1}
\end{center}
\end{figure}

In the rf gate reflectometry experiments shown in Figs.~\ref{fig:Figure1},~\ref{fig:Figure2}, and~\ref{fig:Figure3}, a driving frequency of 583 MHz is used, which is close to the resonant frequency and ideal to distinguish the acceptor-lead and inter-acceptor transitions. The full frequency response of the reflected signal from the LC-circuit reveals more information about the tunnel rates involved in the experiment~\cite{Hile:2015}. Frequency domain measurements of the amplitude and phase response around the resonance of the LC circuit are shown in Fig.~\ref{fig:FigureS1}. These measurements show that the acceptor-lead signal mainly changes the amplitude of the reflected signal, while the inter-acceptor signal mainly reduces the resonant frequency.

This difference in response of the LC-circuit can be explained by a difference in tunnel rate between these two processes~\cite{House:2015}. When the tunnel rate is much faster than the resonant frequency ($\Gamma \gg \omega$), the tunnel process will predominantly add capacitance to the circuit (dispersive response), which in turn shifts the resonance to a lower frequency. This is observed for the inter-acceptor tunneling process, thereby revealing a tunnel rate much faster than the resonant frequency. When the tunnel rate is of the order of the resonant frequency ($\Gamma \sim \omega$), the tunnel process will predominantly add resistance to the circuit (dissipative response). This added resistance can either bring the circuit closer or further away from its ideal matching condition, depending on the initial impedance of the circuit, thereby either reducing or increasing the reflection amplitude. A reduction in reflection amplitude is observed for the acceptor-lead tunneling process, thereby revealing a tunnel rate of the order of the resonant frequency.

\begin{figure}[t]
\begin{center}
\includegraphics[width=88mm]{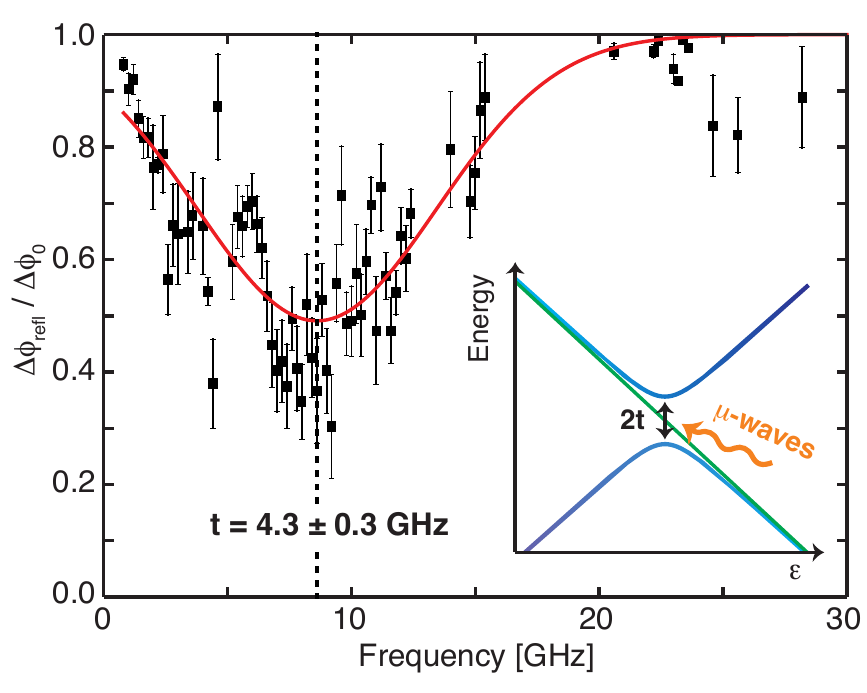}
\caption{Inter-accpetor continuous-wave microwave excitation experiment. The change in phase shift of the inter-acceptor tunnel signal as a function of microwave frequency, using a constant microwave power applied to the drain electrode of the transistor. The resonant dip reveals the tunnel coupling between the atoms measured as $t$ = 4.3$\pm$0.3 GHz. The inset shows a schematic representation depicting the excitation from the bonding to the anti-bonding singlet state with microwaves.}
\label{fig:FigureS2}
\end{center}
\end{figure}

To find the tunnel coupling between the two acceptor atoms, we probe the resonance between the bonding and anti-bonding singlet $S_{HH}$ states, as shown in the inset of Fig.~\ref{fig:FigureS2}, in accordance with a measurement on two electrons on a hybrid donor-quantum dot system~\cite{Urdampilleta:2015}. Here we make use of the bias-tee coupled to the drain lead, as shown in Fig.~\ref{fig:Figure1}a. The inter-acceptor signal is probed with rf gate reflectometry, while continuous wave microwaves are applied to the drain electrode. When the transition between the bonding and anti-bonding $S_{HH}$ states is excited by the applied microwaves, the occupation of the bonding state is lowered, while the occupation of the anti-bonding state is increased. As the bonding and anti-bonding singlet states give rise to an opposite response of the resonator~\cite{Urdampilleta:2015}, the observed $\Delta\phi_\text{refl}$ is expected to decrease when the occupation in the anti-bonding state is increased and completely disappear when the occupations in the system become unpolarised.

In order to keep the input power of the applied microwaves constant during the experiment, the splitting of the acceptor-drain signal is also probed during the experiment. The splitting of this signal has no dependence on the microwave frequency and is solely dependent on the microwave power applied at the drain. Using this measurement we can calibrate for any frequency dependent losses in the system and use the same input power for every frequency. The response of the inter-acceptor tunnel signal is shown in Fig.~\ref{fig:FigureS2} as $\Delta\phi_\text{refl}/\Delta\phi_0$, with $\Delta\phi_0$ being the  signal without microwaves. The resonance dips at a frequency of the 8.6$\pm$0.6 GHz, which demonstrates a tunnel coupling between the atoms of 4.3$\pm$0.3 GHz. The width of this resonance indicates a $T^*_2$ for these charge states of $\sim$ 200 ps, similar to what was found for a two-electron system on a hybrid donor-quantum dot~\cite{Urdampilleta:2015}.

\begin{figure}[b]
\begin{center}
\includegraphics[width=88mm]{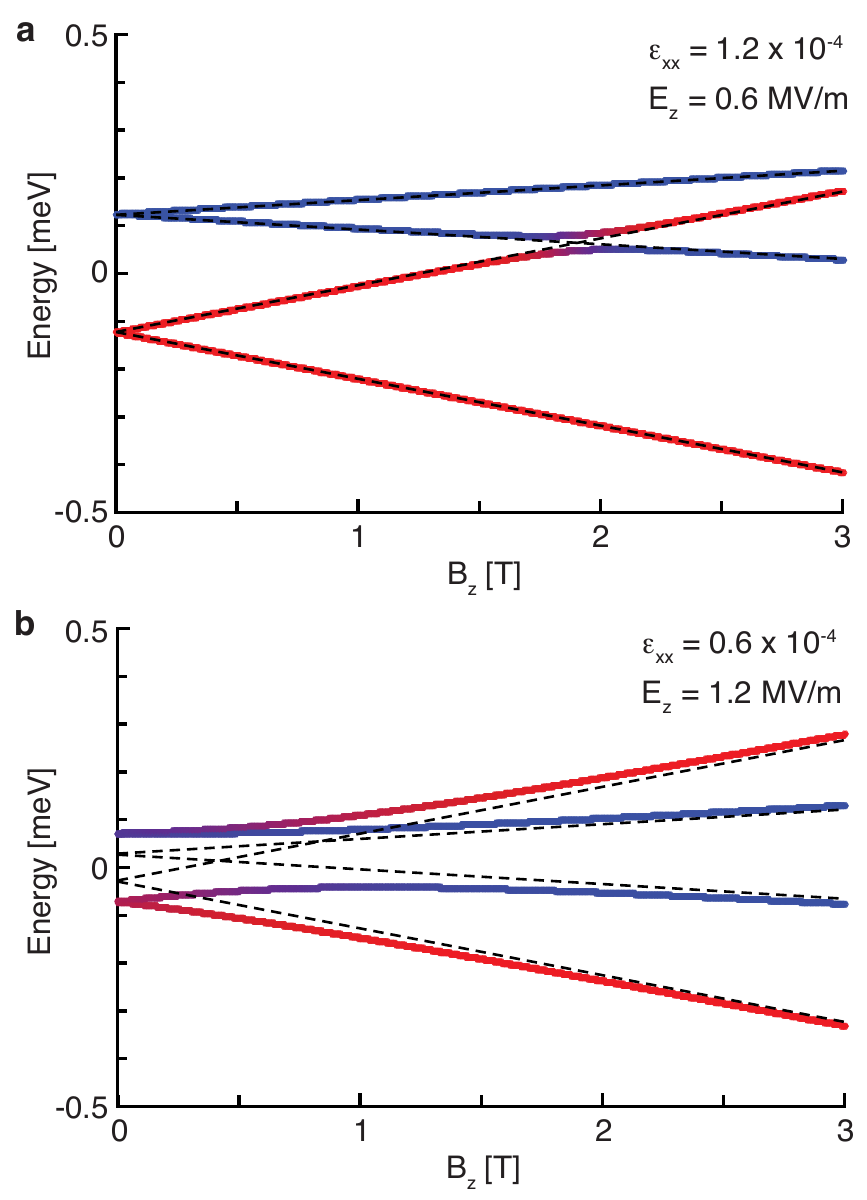}
\caption{Examples of the Zeeman effect on $A^0$-states. \textbf{a}, Magnetic field dependence of the energy levels of a hole bound to a single boron acceptor in silicon with a compressive strain of  1.2$\times$10$^{-4}$ parallel with the magnetic field and an electric field of 0.6 MV/m perpendicular to the magnetic field. \textbf{b}, Magnetic field dependence of the energy levels of a hole bound to a single boron acceptor in silicon with a compressive strain of  0.6$\times$10$^{-4}$ parallel with the magnetic field and an electric field of 1.2 MV/m perpendicular to the magnetic field. Dashed lines show the linear projection of the high magnetic field Zeeman effect, showing a change in $\Delta_\text{LH}$ in \textbf{b}.}
\label{fig:FigureS3}
\end{center}
\end{figure}

\subsection{Two-hole states}

In bulk silicon the single hole ground state of an acceptor atom is fourfold degenerate, consisting of heavy hole and light hole states. The effect of strain, electric field and magnetic field on these states is well described by Bir et al.~\cite{Bir:1963.1, Bir:1963.2}. Strain and electric field split the heavy and light hole manifolds, leading to a splitting of $\Delta_\text{LH}$ at zero magnetic field. A magnetic field splits all four spin-states, where this Zeeman effect strongly depends on the local environment of the atom. The two atoms probed in the experiments are found to have heavy hole ground states. The $\Delta_\text{LH}$ is assumed to be different for the two acceptors, as they experience a different local environment. As an example, the Zeeman effect is calculated for two different local environments where the heavy holes form the ground state at zero magnetic field, shown in Fig.~\ref{fig:FigureS3}. For the first example we consider the situation where a compressive strain of 1.2$\times$10$^{-4}$ is present parallel to the magnetic field and an electric field of 0.6 MV/m is present perpendicular to the magnetic field, see Fig.~\ref{fig:FigureS3}a. In Fig.~\ref{fig:FigureS3}b the situation with half of the strain and double the electric field is considered. In this second example, we note that if a linear approximation is taken for the hole-spin states in the high magnetic field region and traced back to zero magnetic field (dashed lines), a different heavy-light hole splitting is found than at zero magnetic field. This effect, caused by a rotation of the quantization axis of the angular momentum of the hole, has been observed in our experiment presented in Fig.~\ref{fig:Figure4}.

\begin{figure}[t]
\begin{center}
\includegraphics[width=88mm]{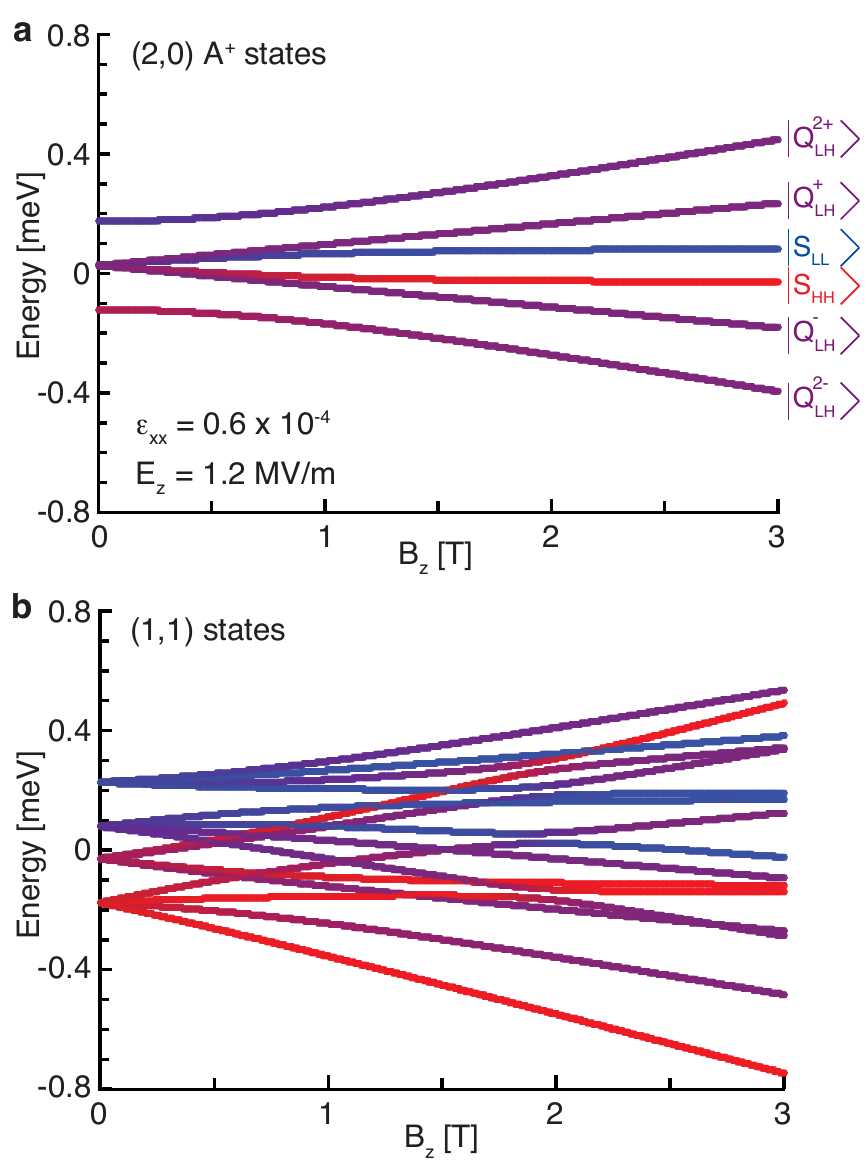}
\caption{Examples of the Zeeman effect on two-hole states. The light or heavy hole character of each state is represented by the line color, where red represents heavy hole character and blue represents light hole character. \textbf{a}, A$^{+}$ energies for the same local environment as in Fig.~\ref{fig:FigureS3}b and $J_\text{LH}=0$. A clear difference in splitting between the S$_\text{HH}$ and S$_\text{LL}$ states in the low and high magnetic field limit is visible. \textbf{b}, The (1,1)-configuration product states of the two single hole states described in Fig.\ref{fig:FigureS3}a and \ref{fig:FigureS3}b.}
\label{fig:FigureS4}
\end{center}
\end{figure}

For the doubly occupied and positively charged acceptor state (A$^{+}$) the exchange energy between two holes from the same manifold far exceeds the heavy-light hole splitting ($J_\text{HH},J_\text{LL} \gg \Delta_\text{LH}$). Due to this exchange interaction, only 6 out of the 16 possible two-hole states can be bound to a boron atom in silicon~\cite{Smit:2004}. A similar sixfold manifold is found for two-holes bound to two closely placed boron atoms~\cite{Salfi:2016.1} and the neutral states of two holes bound to group II acceptors~\cite{Bhattacharjee:1972, Kartheuser:1973}. As these six states are build from combinations of different spin states, one heavy hole singlet state, one light hole singlet state and four light/heavy-quadruplet states are found, which we will write as:
\begin{flalign}
&\mid\text{S}_\text{HH}\rangle = 1/\sqrt{2} \hspace{1.5mm} \left(\mid \Uparrow\Downarrow\rangle - \mid \Downarrow\Uparrow\rangle\right),\\
&\mid\text{S}_\text{LL}\rangle = 1/\sqrt{2} \hspace{1.5mm}\left(\mid \uparrow\downarrow\rangle - \mid \downarrow\uparrow\rangle\right),\\
&\mid\text{Q}^{2-}_\text{LH}\rangle = 1/\sqrt{2} \hspace{1.5mm} \left(\mid \Downarrow\downarrow\rangle - \mid \downarrow\Downarrow\rangle\right),\\
&\mid\text{Q}^-_\text{LH}\rangle = 1/\sqrt{2} \hspace{1.5mm} \left(\mid \Downarrow\uparrow\rangle - \mid \uparrow\Downarrow\rangle\right),\\
&\mid\text{Q}^+_\text{LH}\rangle = 1/\sqrt{2} \hspace{1.5mm} \left(\mid \Uparrow\downarrow\rangle - \mid \downarrow\Uparrow\rangle\right),\\
&\mid\text{Q}^{2+}_\text{LH}\rangle = 1/\sqrt{2} \hspace{1.5mm} \left(\mid \Uparrow\uparrow\rangle - \mid \uparrow\Uparrow\rangle\right),
\end{flalign}
where $\Uparrow, \Downarrow$ represent the heavy hole states and $\uparrow, \downarrow$ represent the light hole states. Furthermore, when an exchange interaction between the heavy and light holes is introduced, $J_\text{LH}$, the Q$_\text{LH}$ state splits into a singlet and triplet state~\cite{Salfi:2016.3}. In Fig.~\ref{fig:FigureS4}a an example of the magnetic field dependence of the A$^+$-states of an acceptor is shown, using the A$^0$ states from Fig.~\ref{fig:FigureS3}b and neglecting any exchange between the light and heavy holes ($J_\text{LH}$ = 0). The colors in the figure show the heavy-hole (red) and light-hole (blue) character of these states, where the states built from one heavy and one light hole turn purple. The splitting between the heavy hole singlet and the light hole singlet ($E_{\text{S}_\text{LL}}$ - $E_{\text{S}_\text{HH}}$ = 2$\Delta_\text{LH}$) is found to be different in the high magnetic field limit compared to the zero magnetic field splitting, in line with was found in Fig.~\ref{fig:FigureS3}b.

In the situation where each hole is bound to a different acceptor, the two hole state can be described as a product state of the single particle states, leading to an energy spectrum of 16 states divided into 4 fourfold degenerate manifolds at zero magnetic field. Each fourfold manifold splits into a singlet and a triplet state when an exchange interaction is introduced. In our experiments the heavy-hole manifold has the lowest energy at zero magnetic field and can be described as:
\begin{flalign}
&\mid \text{S}_\text{HH}\rangle = 1/\sqrt{2} \hspace{1.5mm}\left(\mid \Uparrow,\Downarrow\rangle - \mid \Downarrow,\Uparrow\rangle\right),\\
&\mid \text{T}^-_\text{HH}\rangle = \mid \Downarrow,\Downarrow\rangle,\\
&\mid \text{T}^0_\text{HH}\rangle = 1/\sqrt{2} \hspace{1.5mm} \left(\mid \Uparrow,\Downarrow\rangle + \mid \Downarrow,\Uparrow\rangle\right),\\
&\mid \text{T}^+_\text{HH}\rangle = \mid \Uparrow,\Uparrow\rangle.
\end{flalign}
The energy splitting to the next two manifolds is given by the $\Delta_\text{LH}$ of each acceptor, where these two-hole states consist of one heavy and one light hole state. Finally, the light-hole manifold is split from the heavy-hole manifold by the sum of both $\Delta_\text{LH}$'s. In Fig.~\ref{fig:FigureS4}b an example of such a (1,1) level configuration is shown for a combination of the A$^0$-states presented in Fig.~\ref{fig:FigureS3}a and~\ref{fig:FigureS3}b. Although many level crossings are observed, we highlight the fact that in this situation where the heavy holes form the ground state manifold, the ground state in the (1,1) region remains the T$^-_\text{HH}$ state for any positive magnetic field.

In the transport measurements shown in Fig.\ref{fig:Figure4} the (1,1)$\rightarrow$(2,0) transition energies at zero magnetic field and as a function of magnetic fields up to 3 T are probed. In these experiments, a voltage of +10 mV is applied to the drain, while the current is measured at the source. This bias direction gives a transport direction for the holes from the drain to the source. From the rf gate reflectometry experiments, as shown in Fig.~\ref{fig:Figure2}, it is deduced that the acceptor with a charge transition from 0 to 1 hole has a stronger tunnel coupling to the drain (detectable with rf reflectometry) than the acceptor with a charge transition from 1 to 2 holes (not detectable with rf reflectometry). It follows that the probed sequential tunnelling involves the (1,1)$\rightarrow$(2,0) transition, where the two observed bias triangles are formed by the two following hole charge transfer processes:
\begin{flalign}
(1,0) \rightarrow (1,1) \rightarrow (2,0) \rightarrow (1,0),\\
(2,1) \rightarrow (1,1) \rightarrow (2,0) \rightarrow (2,1).
\end{flalign}
The current measured within the bias triangle region is around 5 pA at the baseline at zero magnetic field and declining to the measurement noise floor of $\sim$ 100 fA at the other side of the triangle, indicating a total tunnel rate of the sequential tunnel process around 30 MHz and lower. We recall that the resonant tunnel-rate between the (1,1) and (2,0) S$_\text{HH}$ states was measured at 4.3 GHz and the tunnel-rate from the drain to the acceptor was found to be around the resonant frequency of 583 MHz, indicating that the tunnel-rate between the second acceptor and the source is the limiting tunnel process. However, we note that due to off-resonant tunneling or tunneling between two states consisting of different spins, the inter-acceptor tunnel rate can be strongly suppressed, thereby possibly becoming the limiting tunnel process. Furthermore, the ratio between this inter-acceptor tunnel rate and the relaxation rate of each excited state in the (1,1)-configuration determines if this excited state will have any effect on the measured current. The transport measurement presented in Fig.~\ref{fig:Figure4} can be explained by only taking the heavy-hole manifold of the (1,1)-configuration into account. This indicates that any state which includes any light-hole has a fast relaxation to the heavy hole ground state, which matches with the fast light-to-heavy hole relaxation process observed as a relaxation hotspot in Fig.~\ref{fig:Figure3}.

\subsection{Pauli spin blockade in rf gate reflectometry}

The shift in phase response of the reflected signal as shown in Fig.~\ref{fig:Figure2}b is ascribed to the addition of a quantum capacitance ($C_\text{Q}$) to the LC-circuit~\cite{Petersson:2010, Colless:2013, Verduijn:2014, Gonzalez-Zalba:2015, House:2015} when holes can tunnel between the two atoms, given by:
\begin{equation}
		C_\text{Q}  = \frac{1}{2} \Delta\alpha_\text{TG}^2 q^2 4t^2\left(\epsilon^2 + 2t^2 \right)^{-3/2},
\label{eq:EQM1}
\end{equation}
where $\Delta\alpha_\text{TG}$ is the difference in capacitive coupling from the top gate to each acceptor, found to be 0.14 eV/V in the transport measurements, $q$ is the electron charge, $\epsilon$ is the energy detuning between the two acceptors, and $t$ is the tunnel coupling between the (1,1) and (2,0) S$_\text{HH}$ states. All other tunnel couplings between (1,1) and (2,0) states are considered to be much smaller and therefore not contributing to a change in the detected reflected signal. The total phase shift ($\Delta \phi$) is determined by the difference in occupation, given by the Boltzmann distribution, of the lower and upper branch of the heavy hole singlet state:
\begin{equation}
		\Delta \phi = C_\text{Q} \frac{e^{\left(-E_{\text{S}^-}/k\text{T}\right)} - e^{\left(-E_{\text{S}^+}/k\text{T}\right)}}{\sum e^{\left(-E_{i}/k\text{T}\right)}},
\label{eq:EQM2}
\end{equation}
where $E_{\text{S}_-}$ and $E_{\text{S}_+}$ are the energies of the bonding and anti-bonding singlet states respectively and $E_{i}$ describes the energies of all the involved states. Using Eq.~\ref{eq:EQM1} and~\ref{eq:EQM2}, the magnetic field dependence of the phase shift caused by the inter-acceptor tunneling is calculated as shown in Fig.~\ref{fig:Figure2}c. We note that although the crossing of the Q$_\text{LH}$ state and the S$_\text{HH}$ state has some influence on the Boltzmann distribution of the S$_\text{HH}$ states, this influence is not distinguishable in the measurement. However, for completeness, we have displayed the $B_{\text{S}\rightarrow\text{Q}}$ crossing point in Fig.~\ref{fig:Figure2}c, as found in the relaxation-hotspot measurement together with an estimated g-factor of the Q$_\text{LH}$ state.

\begin{figure}
\begin{center}
\includegraphics[width=88mm]{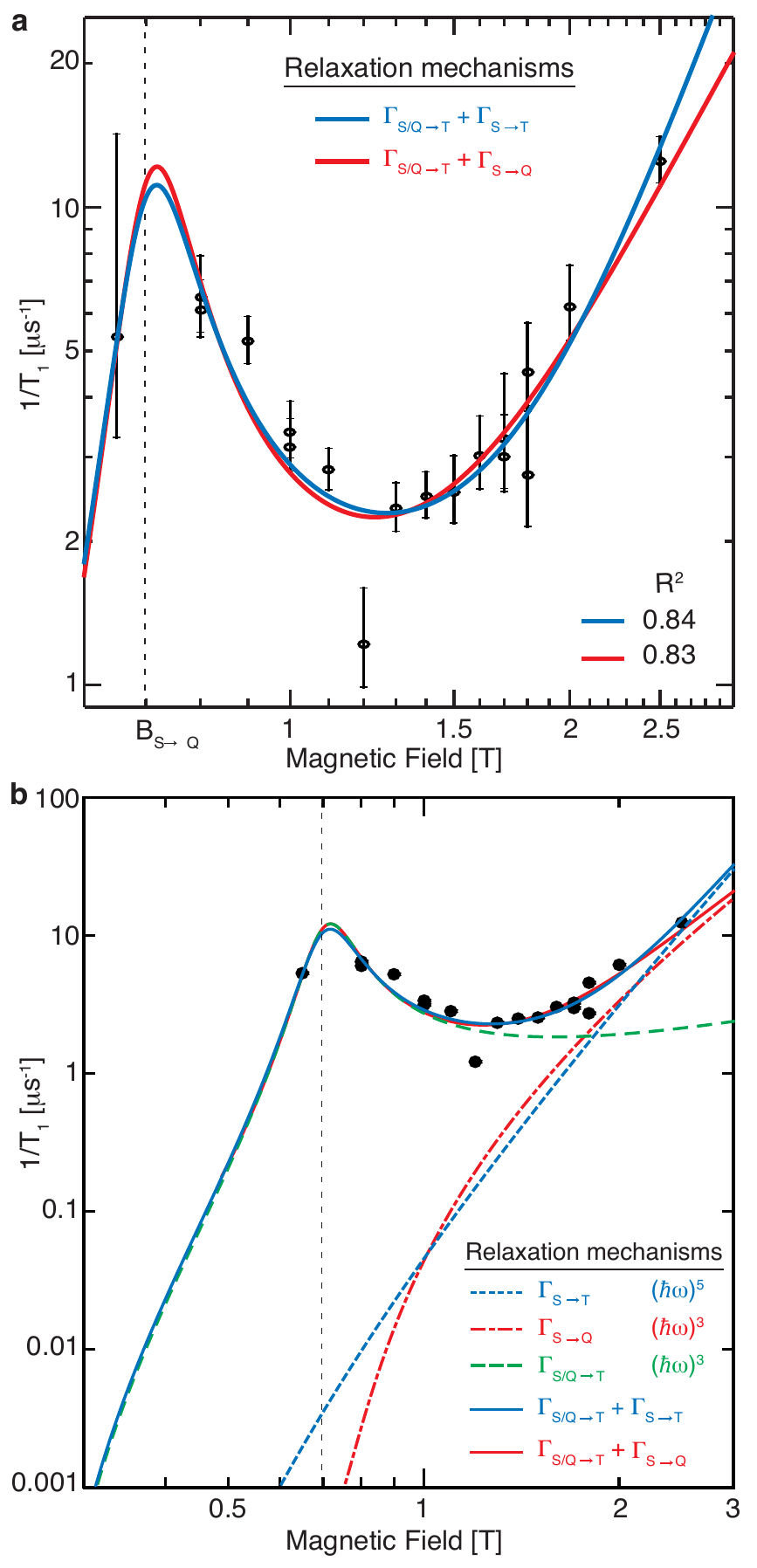}
\caption{Relaxation rate models. \textbf{a}, Comparison between the fit of the relaxation rate data with a model combining the hotspot-relaxation mechanism with either the S$_\text{HH}$ to T$^-_\text{HH}$ relaxtion (blue) or with the S$_\text{HH}$ to Q$^{2-}_\text{LH}$ relaxation (red). Both fits have a comparable value for $R^2$. \textbf{b}, Comparison of the modeled relaxation mechanisms on a larger scale of magnetic field and $T_1$.}
\label{fig:FigureS5}
\end{center}
\end{figure}

\subsection{Relaxation mechanisms}

At magnetic fields much higher than the $B_{\text{S}\rightarrow \text{Q}}$ mixing point the dominant relaxation mechanism comes from either the relaxation to the (1,1) T$^-_\text{HH}$ state or the relaxation to the (2,0) Q$^{2-}_\text{LH}$ state, given by Eq.~\ref{eq:EQ1}. In Fig.~\ref{fig:FigureS5}a the fits of the relaxation rate data with a combination of Eqs.~\ref{eq:EQ1} and~\ref{eq:EQ2} is given for the relaxation process to the T$^-_\text{HH}$ state (blue) and the Q$^{2-}_\text{LH}$ state (red) are shown. Our data is insufficient to determine if a $(\hbar\omega)^3$ or $(\hbar\omega)^5$ relaxation mechanism is dominant at high magnetic fields. The relaxation amplitudes extracted from these two fits, which are interpreted as the relaxation time at a field of 1 T relative to the corresponding level-crossing, is found to be 0.5 $\mu$s for $A_\text{SQ}$ and 5 $\mu$s for $A_\text{ST}$. Both of these values are in good agreement with the theoretical predictions for light-heavy hole relaxation ($T_1$ = 0.14 $\mu$s at 1T~\cite{Ruskov:2013}) and heavy-heavy hole relaxation ($T_1$=40 $\mu$s at 0.5T~\cite{Salfi:2016.2}). Fig.~\ref{fig:FigureS5}b shows the extracted relaxation rates on a larger scale of $1/T_1$, demonstrating the more than 3 orders of magnitude increase in relaxation rate at the hotspot. Furthermore, it reveals that if this hotspot behavior can be moved to a higher magnetic field, by increasing the heavy-light hole splitting or decreasing the g-factor, relaxation times longer than 1 ms could be achieved at magnetic fields around 0.5T.

\subsection{Relaxation hotspot}

The observed relaxation hotspot is explained by the mixing of the bonding S$_\text{HH}$ state and the (2,0) Q$_\text{LH}^{2-}$ state, which allows a light-to-heavy hole relaxation to the (1,1) T$_\text{HH}^-$ state. This relaxation process is expected to have a $(\hbar\omega)^3$ dependence and furthermore depends linearly on the light hole character of the mixed state. Here we show the calculation of the light hole character of the probed state.

Around the anti-crossing we can write the energies of the lower (-) and upper (+) states as:
\begin{equation}
		E_\pm = \frac{\Delta E}{2}\pm\frac{1}{2}\sqrt{(\Delta E)^2+\left(2 t_\text{SQ}\right)^2},
\label{eq:EQM3}
\end{equation}
where $\Delta E$ is the energy splitting between the states in the absence of mixing ($E_{\text{Q}_\text{LH}^{2-}}$ - $E_{\text{S}_\text{HH}}$), $E_{\text{S}_\text{HH}}$ is taken as the reference energy ($E=0$). We define the amount of light hole character of each state as the chance to measure it as the Q$_\text{LH}$ state in the basis of Q$_\text{LH}$ and S$_\text{HH}$, written as $\xi_\pm$ and given by:
\begin{equation}
		\xi_\pm = \frac{1}{2}\mp\frac{\Delta E}{2\sqrt{(\Delta E)^2+\left(2 t_{SQ}\right)^2}}.
\label{eq:EQM4}
\end{equation}
The relaxation measurement is performed at zero detuning, which is the original position of $E_{\text{S}_\text{HH}}$ in the absence of mixing. The state probed here can be written as a combination of the + and - states, depending on their relative energy difference with the measurement point $\left(\frac{\lvert E_\pm\rvert}{\lvert E_-\rvert+\lvert E_+\rvert}\right)$. We note that $\lvert E_-\rvert \xi_-$ = $\lvert E_+\rvert \xi_+$ = $t_\text{SQ}^2/\sqrt{(\Delta E)^2+\left(2 t_{SQ}\right)^2}$ and also that $(\lvert E_-\rvert+\lvert E_+\rvert)$ is equal to $\sqrt{(\Delta E)^2+\left(2 t_\text{SQ}\right)^2}$, which gives the total light hole character at the zero energy point as:
\begin{equation}
		\xi_{E=0} = \frac{\lvert E_-\rvert \xi_- + \lvert E_+\rvert \xi_+}{\lvert E_-\rvert+\lvert E_+\rvert} = \frac{2t_\text{SQ}^2}{(\Delta E)^2+\left(2 t_\text{SQ}\right)^2},
\label{eq:EQM5}
\end{equation}
as used in Eq.~\ref{eq:EQ2}.

\subsection{Acceptor-based spin-orbit qubits}

The key properties of acceptor-based spin-orbit qubits, such as the dipole coupling and decoherence time between the chosen qubit states, are predicted to strongly depend on changes to the symmetry of the silicon crystal and the confinement of the hole~\cite{Bir:1963.1, Bir:1963.2, Ruskov:2013, Salfi:2016.2, Salfi:2016.3, Abadillo-Uriel:2016}. Unstrained silicon ensemble measurements have established an electric dipole coupling of 0.26 Debye~\cite{Kopf:1992} and relaxation and coherence times in the micro-second regime~\cite{Song:2011}. These acceptor systems are promising for fast electrically-driven spin-orbit qubits with a strong coupling to cavities and phonons~\cite{Ruskov:2013}, although a longer coherence time would be desirable. An optimal regime, where the coherence and relaxation times are prolonged while the strong electric dipole coupling is maintained, has been predicted in the small strain regime, giving a small splitting and strong mixing between the heavy and light hole states~\cite{Salfi:2016.3}. The possibility to lengthen the acceptor relaxation time has been demonstrated in experiments on intentionally heavily strained bulk samples, where $T_1$ times up to 100 ms are measured~\cite{Ludwig:1962, Dirksen:1989}. The access to and control over single atoms, as has been demonstrated here, is necessary to probe the physics in an environment with a small local strain, which is the crucial regime for the realisation of acceptor qubits.

\section*{Acknowledgments} 
The device has been designed and fabricated by the TOLOP Project partners, see http://www.tolop.eu. This work was supported by the ARC Centre of Excellence for Quantum Computation and Communication Technology (CE110001027) and the ARC Discovery Project (DP120101825).

\bibliographystyle{naturemag-doi}

\end{document}